# 4T2R X-ReRAM CiM Array for Variation-tolerant, Low-power, Massively Parallel MAC Operation

Fuyuki Kihara[1*], Seiji Uenohara[3], Satoshi Awamura[3], Naoko Misawa[1], Chihiro Matsui[1] and Ken Takeuchi[1,2]
[1] Dept. of Electrical Engineering and Information Systems, The University of Tokyo, 7-3-1 Hongo, Bunkyo-ku, Tokyo 113-8656, Japan
[2] Systems Design Lab., Graduate School of Engineering, The University of Tokyo, 2-11-16 Yayoi, Bunkyo-ku, Tokyo 113-0032, Japan
[3] Nuvoton Technology Corporation Japan, 1 Kotari-yakemachi, Nagaokakyo City, Kyoto 617-8520, Japan
*E-mail: kihara@co-design.t.u-tokyo.ac.jp

*Abstract*— Computation-in-Memory (CiM) is attracting attention as a technology that can perform MAC calculations required for AI accelerators, at high speed with low power consumption. However, there is a problem regarding power consumption and device-derived errors that increase as row parallelism increases. In this paper, a 4T2R ReRAM cell and an 8T SRAM CiM suitable for CiM is proposed. It is shown that adopting the proposed 4T2R ReRAM cell reduces the errors due to variation in ReRAM devices compared to conventional 4T4R ReRAM cells.

## I. Introduction

Computation-in-Memory (CiM) is considered one of the methods to overcome the von Neumann bottleneck and can perform simple calculations such as multiply-accumulate (MAC) operations with high parallelism, low power consumption, and low precision, and is effective for algorithms that include a lot of matrix and vector calculations such as neural network and signal processing [1]. In particular, CiM using non-volatile memory (NVM) such as ReRAM is attracting attention as a possible implementation for AI edge devices [2-4].

ReRAM and SRAM each have different characteristics, and they are used depending on the situation. ReRAM is non-volatile and easy to increase the capacity, but the device-derived variation is large. SRAM is volatile and has a large cell area, but it has a low rewrite cost and is faster. For AI accelerators, it is thought to be effective to use SRAM CiM for layers that are rewritten frequently, and ReRAM CiM for layers that require large capacity. For example, in a Transformer [5], there is little rewriting in the fully-connected (FC) layer, but there is a lot of rewriting in the self-attention (SA) layer (Fig. 1(a)). Therefore, it has been proposed to apply ReRAM CiM to the FC layer and SRAM CiM to the SA layer. The structure and operation of the ReRAM are shown in Fig. 2(a). ReRAM is composed of metal oxides such as $TaO_X$, and the resistance value is changed by the movement of oxygen ions due to an electric field. Fig. 2(b) shows the measured conductance of the ReRAM. Although multi-level writing and reading are possible, the conductance has a variation of over 50%.

In a CiM array, as row parallelism increases, the current flow increases, leading to an increase in power consumption and a decrease in dynamic range. Current Limiting Differential Readout (CuLD) [6] has been proposed as a technology to prevent this.

## II. Current Limiting Differential Readout (CuLD)

### A. Principle of CuLD circuit

CuLD circuit [6] is one of the CiM architectures that is considered to not increase power consumption even when the row parallelism is large. Fig. 1(b) shows the principle circuit of CuLD. In this circuit, four 1T1R cells are integrated to form a 4T4R cell (Fig. 3), representing a single coded weight. The ReRAM in the upper left and lower right, and the ReRAM in the upper right and lower left each have the same resistance values, $R_p$ and $R_n$. When WL is active, the two ReRAMs above are enabled, and when WLB is active, the two ReRAMs below are enabled. Let the current flowing through the ReRAM with resistance values $R_p$ and $R_n$ is set to $I_p$ and $I_n$, the current flowing through BL will change between $I_p$ and $I_n$ depending on whether WL or WLB is active. WL and WLB are input with complementary PWM signals whose pulse width ($X_i$) determined by the input value. Fig. 3 also shows the operation of CuLD when performing a MAC operation. The slope of $V_{xp}$ and $V_{xn}$ changes depending on the input waveform. The amount of charge flowing through the 4T4R cell in the $i$-th row during the WL or WLB operating time ($X_{max}$) is expressed as follows:

$$Q_{p,i} = I_{p,i} X_i + I_{n,i}(X_i - X_{max}) \quad (1)$$

$$Q_{n,i} = I_{p,i}(X_i - X_{max}) + I_{n,i} X_i \quad (2)$$

where $Q_{p,i}$ and $Q_{n,i}$ are the amounts of electric charge flowing through BL and BLB from $i$-th cell, respectively. The current flowing through BL and BLB is replicated by the current mirror circuit and charged to the capacitor $C_p$ and $C_n$. As a result, the potential difference between the two capacitors is as follows:

$$V_x \equiv V_{xp} - V_{xn} = \frac{1}{C}\sum_i (2X_i - X_{max})(I_{p,i} - I_{n,i}) \quad (3)$$

where $C$ is capacitance of $C_p$ and $C_n$ in Fig. 1(b). Here, the first term on the right-hand side of equation (3) represents input value, and the second term represents the weight. Therefore, the output voltage $V_x$ is the result of the MAC operation of signed inputs and weights.

The equivalent circuit focusing on the ReRAM array during the time $k$ WLs or WLBs are active is shown in Fig. 4. When the parallel composite resistance value ($R_p // R_n$) is a constant value in all the rows, the current flowing through the cell becomes a constant value ($= I_{BIAS}/k$). If this condition is violated, there is a risk of errors in the results. When considering the parallel composite resistance value, it is the one with the smaller resistance value of the two ReRAMs that has a stronger impact, so it is necessary to pay attention to the variation in the lower resistance value of $R_p$ and $R_n$. In particular, if there is a mismatch in the resistance values within the same 4T4R cell, equations (1) and (2) will no longer be valid.

### B. Relationship between weight and ReRAM value

When weight is set to +1, the Low Resistance State (LRS) resistance is assigned to $R_p$ and the High Resistance State

(HRS) resistance value is assigned to $R_\text{n}$. When a weight is set to -1, the assignment is reversed.

If the weights $a_i$ is to be set to $a_i \in [-1, 1]$, the resistance values of the ReRAM are set as below:

$$R_{\text{p},i} = \frac{2R_\text{HRS}}{R_\text{HRS} + R_\text{LRS} + a_i(R_\text{HRS} - R_\text{LRS})} R_\text{LRS} \quad (4)$$

$$R_{\text{n},i} = \frac{2R_\text{HRS}}{R_\text{HRS} + R_\text{LRS} - a_i(R_\text{HRS} - R_\text{LRS})} R_\text{LRS} \quad (5)$$

where $R_\text{HRS}$ is the highest resistance and $R_\text{LRS}$ is the lowest resistance that ReRAM can represent. In this case, conditions $R_\text{p}//R_\text{n} = \text{const.}$ and $I_{\text{p},i} - I_{\text{n},i} \propto a_i$ are satisfied. In particular, when the weight is 0, the required resistance value is $2R_\text{LRS}$. ReRAM devices used are required to have low resistance variation in the low resistance range.

## III. Proposed 4T2R ReRAM Cell and 8T SRAM Cell

### A. Proposed 4T2R ReRAM Cell

Fig. 5(b) shows the proposed 4T2R ReRAM Cell. This is constructed from four access FETs and two ReRAMs. One end of the two ReRAMs is connected to a common Source Line. The other end is connected to two Bit Lines (BL/BLB) by four FETs driven by two Word Lines (WL/WLB). The FETs connected to the WL connect the left ReRAM to the left BL and the right ReRAM to the right BLB. The FETs connected to WLB connect the left ReRAM to the right BL and the right ReRAM to the left BLB. This allows the current path to be switched by changing the WL/WLB that is activated. The operation of the 4T2R cell as seen from the external circuit of the cell is the same as that of the 4T4R cell.

The write and read operations of the 4T2R cell are shown in Fig. 6. Splitting the Source Line into two, SL and SLB, is effective during ReRAM forming and write operations, but they are connected and made to have the same potential during CiM operations.

As shown in Fig. 7, unlike 4T4R cell, since there are no pairs of ReRAMs in one 4T2R cell that should have the same value, there is no possibility of a mismatch in the resistance value within the cell.

### B. Proposed 8T SRAM Cell

Fig. 5(c) shows proposed 8T SRAM Cell. 8T SRAM is composed of the general 6T SRAM and two access transistors that are activated by WLB. Similarly to the 4T2R cell, it operates by inputting complementary PWM signals to WL and WLB. To prevent unintended writes from occurring while operating as CiM, the WL and WLB voltages are limited to values lower than those used for writing operations.

## IV. Circuit Simulation Results

Parameters used in HSPICE simulations are shown in TABLE I.

### A. 4T2R ReRAM

Fig. 8 shows the simulation results for MAC operations using CuLD with three types of cells: 4T4R cells without mismatch, 4T4R cells with mismatch, and 4T2R cells. The resistance values of the ReRAMs contained in each cell are shown in Fig. 8(b). As shown in Fig. 8(c), if there is a mismatch in the cell, errors will occur in outputs. On the other hand, when using 4T2R cells, the results are close to those obtained without a mismatch in 4T4R cells.

Simulation results for MAC operations using CuLD with four 4T2R ReRAM cells are shown in Fig. 9. The number of cells in the array is 4, with 5 levels of input values and 2 levels of weights. The straight line in the graph is approximate straight lines obtained by least-squares fitting. The range of output voltage ($V_\text{x}$) is 838 mV, and the RMSE is 7.6 mV. The simulation results shown in Fig. 8(c) indicate that the error is larger than $V_\text{x}$ when there is a mismatch within the cell, and that the MAC value is shifted.

Fig. 10 shows simplified layout representation of 4T2R and 4T4R cells. The area of the 4T2R cell has increased due to the crossing of the wiring. The number of ReRAMs does not affect the cell area.

### B. 8T SRAM

Fig. 11(a) shows the simulation results for MAC operations using CuLD with 8T SRAM when the number of cells($N$) is varied. With the 8T SRAM, it can be observed that the operation is the same as when using the 4T2R cell. The relationship between $N$ and $V_\text{x}$ is shown in Fig. 11(b).

Simulation results for MAC operations using CuLD with four 8T SRAM cells are shown in Fig. 12. As with the 4T2R cell, the MAC value can be calculated (see Fig. 9). The range of Vx is 843 mV, and the RMSE is 6.6 mV. The results are similar to those of the 4T2R cell.

Fig. 13 shows the layout of SRAM cells. The layout of the 8T SRAM shown in Fig. 13(b) is based on the 6T SRAM shown in Fig. 13(a). The number of FETs has increased, but unlike the 4T2R cell, there is no increase in area due to wiring crossings. Fig. 13(c) shows the layout of an array with 8T SRAMs laid out. Die photograph of a CuLD with 8T SRAM array manufactured via ROHM 0.18um process is shown in Fig. 14.

## V. Conclusion

In this paper, a 4T2R ReRAM cell and 8T SRAM cell suitable for low power consumption CiM array are proposed. Circuit simulation results show these memory cells can perform MAC operations. The 4T2R cell solves the problem of resistance value mismatch that occurs in the 4T4R cell. In addition, MAC can be calculated using the same method even with 8T SRAM, and the parallelism of the calculation can be improved.

## Acknowledgment

The authors thank S. Yoneda, S. Ito and N. Hattori of NTCJ. This paper is based on results obtained from a project, JPNP23015, subsidized by New Energy and Industrial Technology Development Organization (NEDO).

## References

[1] T. Xiao *et al*, *Appl. Phys. Rev.,* vol. 7, p. 031301, 2020.
[2] A. Yamada *et al.*, *IRPS*, 2023, pp. 1-6.
[3] R. Mochida *et al.*, *Symp. on VLSI*, 2018, pp. 175-176.
[4] S. Mittal, *Mach. learn. knowl. extr.*, vol. 1, no. 1, pp. 75-114, 2019.
[5] N. Misawa *et al.*, *IMW*, 2024, pp. 45-48.
[6] S. Uenohara *et al.*, *arXiv*, 2025, doi: 10.48550/arXiv.2502.00057.
[7] Q. Liu *et al.*, *ISSCC*, 2020, pp. 500-502.
[8] J.M. Correll et al., *Symp. on VLSI*, 2022, pp. 264–265.
[9] X. Si *et al.*, *IEEE J. Solid-state Circuits*, vol. 55, no. 1, pp. 189-202, 2020.
[10] K. -Y. Chung *et al.*, *IEEE Access,* vol. 12, pp. 24254-24261, 2024.

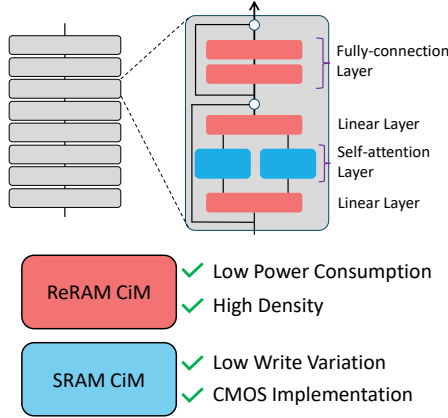
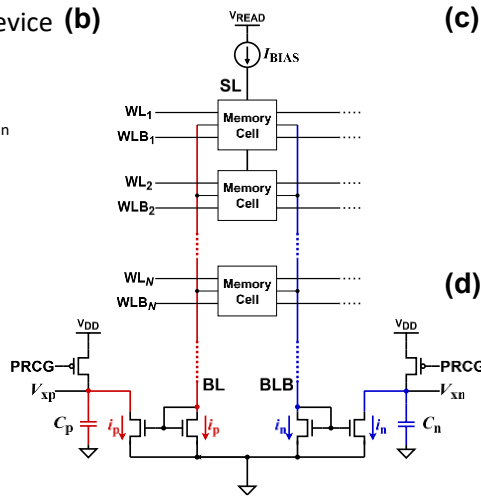
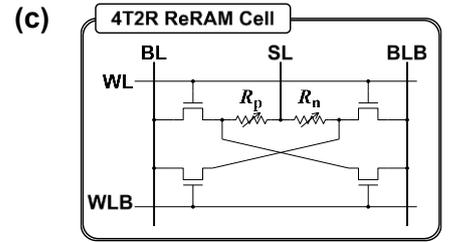
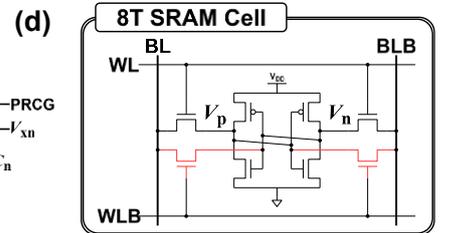

Fig. 1 (a) ReRAM and SRAM CiM in AI accelerator. (b) Principal circuit of CuLD [6]. (c) Circuit of proposed 4T2R ReRAM cell. (d) Circuit of proposed 8T SRAM cell.

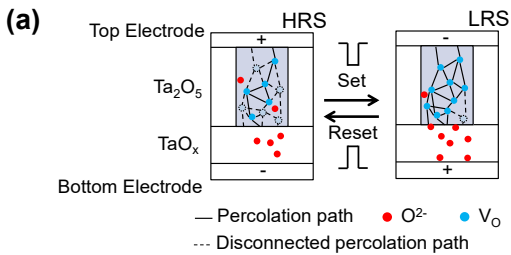
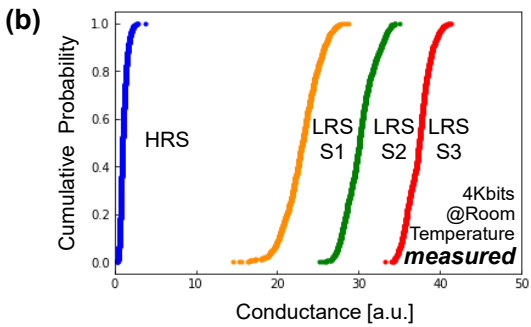

Fig. 2 (a) Switching mechanism of ReRAM [2]. (b) Measured conductance distribution of ReRAM cell.

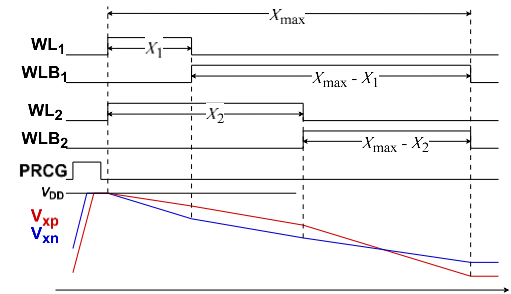
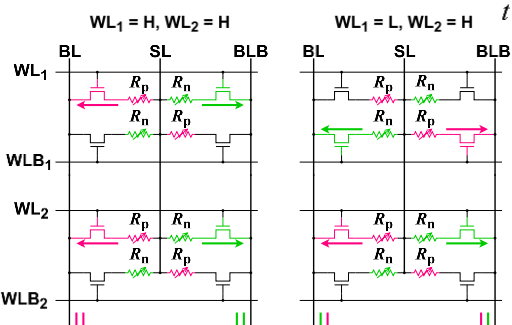

Fig. 3 Example of the operating waveform of CuLD [6]. The current flowing through BLs changes according to the signal of WLs, and the potential of the capacitor changes.

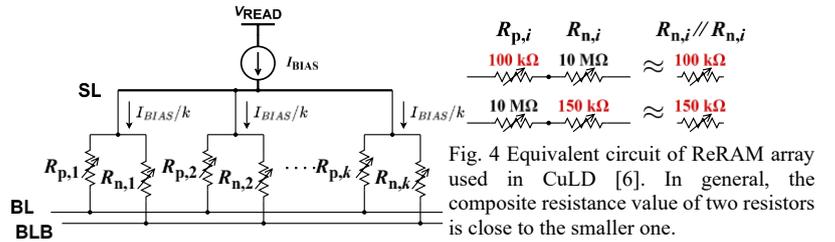

Fig. 4 Equivalent circuit of ReRAM array used in CuLD [6]. In general, the composite resistance value of two resistors is close to the smaller one.

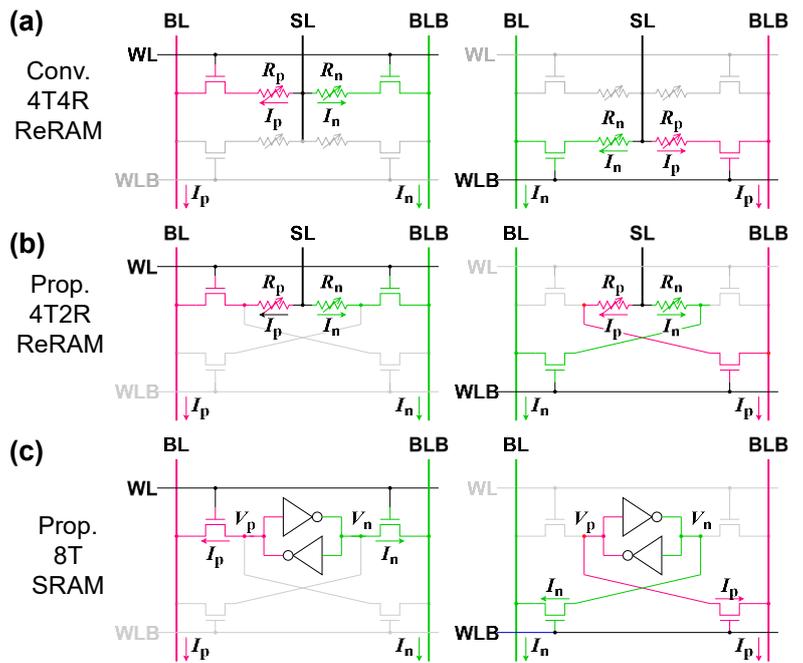

Fig. 5 Schematic of (a) Conventional 4T4R ReRAM (four 1T1R). (b) Proposal 4T2R ReRAM. (c) Proposal 8T SRAM.

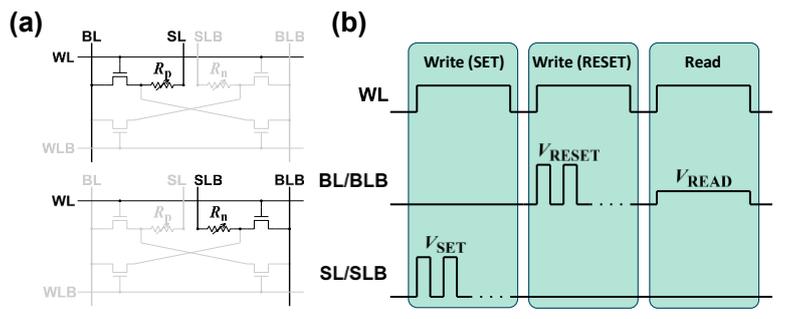

Fig. 6 (a) Circuit and (b) Waveform of write and read operations of 4T2R ReRAM cell. For writing and reading of the single cell, it is recommended that SL and SLB be divided.

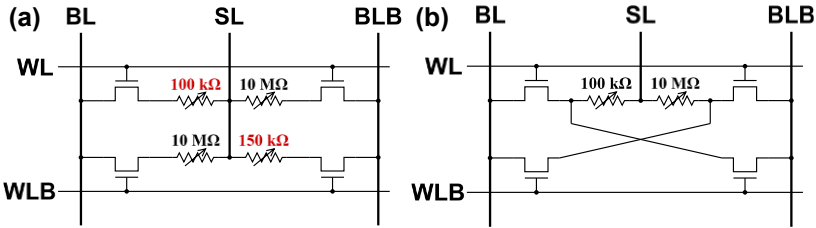

Fig. 7 (a) 4T4R cell with resistance mismatch and (b) 4T2R cell used in the simulation. In the 4T2R cell, no mismatch can occur.

TABLE I. Parameters used in simulations

| | Cell | # of cell $N$ | $I_{BIAS}$ [uA] | $C_p$, $C_n$ [pF] | $V_{DD}$ [V] | $V_{WL}$ [V] | Weight Value | Input Value | WL PWM Width $X_i$ [ns] | Simulation Result |
|---|---|---|---|---|---|---|---|---|---|---|
| i | 4T4R 4T4R w/, w/o mismatch | 32 | 10 | 3 | 0.8 | 0.8 | ±1 | 0, -0.5 | 50, 25 | Fig. 8 |
| ii | 4T2R | 4 | | | | | | ±1, ±0.5, 0 | 0, 25, 50, 75, 100 | Fig. 9 |
| iii | 8T | 2~1024 | | | 1.8 | 0.5 | | 0, -0.5 | 50, 25 | Fig. 11 |
| iv | 8T | 4 | | | | | | ±1, ±0.5, 0 | 0, 25, 50, 75, 100 | Fig. 12 |

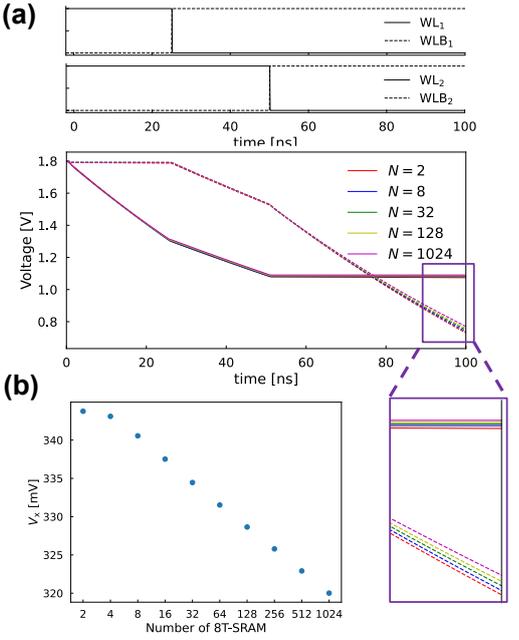

Fig. 11 (a) Waveform example of CuLD with proposed 8T SRAM when number of cells ($N$) is varied. Solid (dashed) lines indicate $V_{xp}$ ($V_{xn}$). (b) $N$-$V_x$ graph.

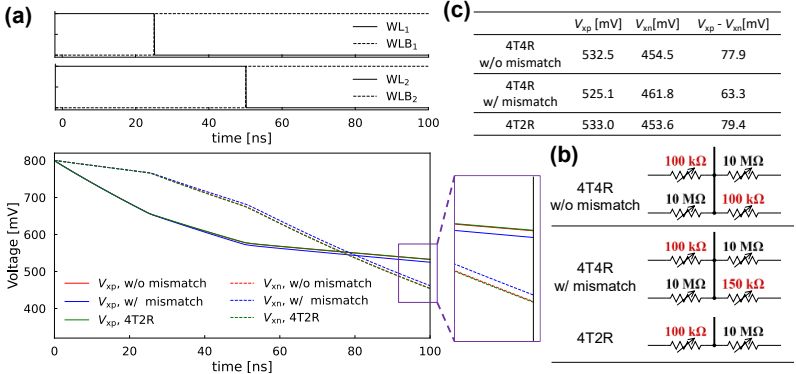

Fig. 8 4T4R cell without mismatch v.s. with mismatch 100k/150k v.s. 4T2R cell. (a) Waveform. (b) Resistance setting. (c) Output voltage.

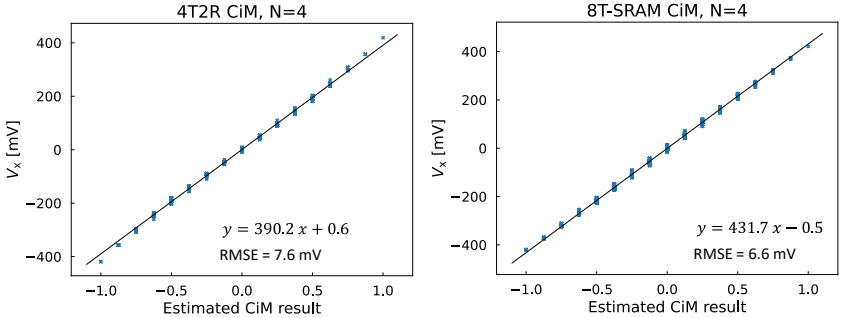

Fig. 9 Four 4T2R cells CiM result.  Fig. 12 Four 8T SRAM cells CiM result.

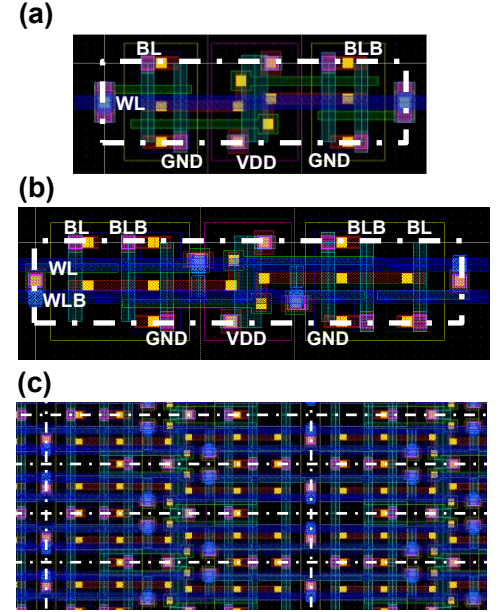

Fig. 13 Layout of (a) one 6T SRAM cell, (b) one 8T SRAM cell, and (c) part of 8T SRAM cell array.

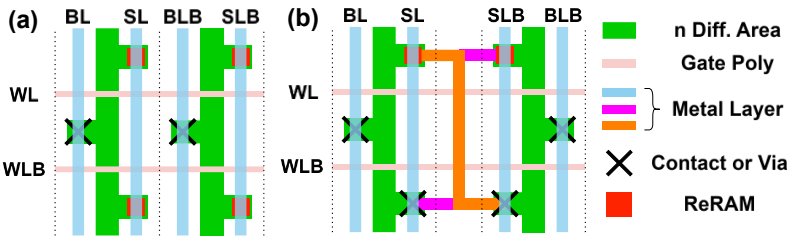

Fig. 10 Stick diagrams of (a) 4T4R cell and (b) 4T2R cell. Intersection of wires causes increase in area.

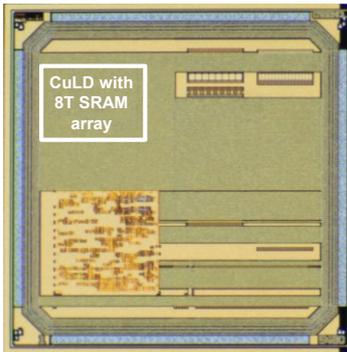

Fig. 14 Die photo of proposed CuLD with 8T SRAM array.

TABLE II. Comparison of CiM Accelerators and CuLD

| | [7] | [8] | [9] | [10] | CuLD[6] | This Work |
|---|---|---|---|---|---|---|
| Input Vector Representation | 1bit Pulse | Analog Voltage | Analog Voltage | PWM, Analog Voltage | PWM | PWM |
| Weight Storage | ReRAM | ReRAM | SRAM | SRAM | ReRAM | ReRAM, SRAM |
| Memory Cell Structure | 1T1R | 1T1R | 8T | 8T | 1T1R | 4T2R |
| Weight Realization | 2 | 2 | 2 | 1 | 4 | 1 |
| # of activated WLs | 1568 | 256 | 50 | 16 | 1024+ | 1024+ |
| # of WLs per Weight | 2 | 1 | 2 | 1 | 2 | 2 |
| # of Effective Inputs | 784 | 256 | 25 | 16 | 512+ | 512+ |
| 1/N Auto Scaling with Current Limit | NO | NO | NO | NO | YES | YES |
| Mismatch tolerance | - | - | - | - | NO | YES |